\documentclass[prb,showpacs,twocolumn,amsmath,amssymb,floatfix,superscriptaddress, longbibliography]{revtex4-1}

\usepackage[utf8]{inputenc}
\usepackage{graphicx}
\usepackage{color}
\usepackage{bbold}
\usepackage{booktabs}
\usepackage{fdsymbol}

\usepackage[T1]{fontenc}
\usepackage[colorlinks,citecolor=blue, urlcolor=blue, linkcolor=blue]{hyperref}
\usepackage{soul}

\graphicspath{{Figures/}}

\newcommand\script[1]{{\fontfamily{pzc}\fontshape{it}\selectfont#1}}

\newcommand\EatDot[1]{}

\newcommand{\DGA}{\text{D}\Gamma\text{A}}
\newcommand{\inverseZ}{\reflectbox{\script{Z}}}

\newcommand{\refs}[1]{Sec.~\ref{sec:#1}}
\newcommand{\refss}[1]{Sec.~\ref{subsec:#1}}

\newcommand{\reff}[1]{Fig.~\ref{fig:#1}}
\newcommand{\reffa}[1]{Fig.~\ref{fig:#1}(a)}
\newcommand{\reffb}[1]{Fig.~\ref{fig:#1}(b)}
\newcommand{\reffc}[1]{Fig.~\ref{fig:#1}(c)}

\newcommand{\refq}[1]{Eq.~(\ref{eq:#1})}
\newcommand{\refqq}[1]{Eq.~(\ref{#1})}

\newcommand{\reft}[1]{Table~\ref{tab:#1}}

\newcommand{\vk}{\mathbf{k}}
\newcommand{\vQ}{\mathbf{Q}}

\newcommand{\wn}{i\omega_n}

\makeatletter
  \def\l@subsubsection#1#2{}%
\makeatother

\begin{document}
	\title{Momentum structure  of the self-energy and its parametrization for the two-dimensional Hubbard model}

\author{P.~Pudleiner}
\affiliation{Institute of Physics, Johannes Gutenberg University, Mainz, Germany}
\email{corresponding author: pudleiner@ifp.tuwien.ac.at}
\affiliation{Institute for Solid State Physics, TU Vienna, Austria}
\author{T.~Sch\"afer}
\affiliation{Institute for Solid State Physics, TU Vienna, Austria}
\author{D.~Rost}
\affiliation{Institute of Physics, Johannes Gutenberg University, Mainz, Germany}
\affiliation{Graduate School Materials Science in Mainz, Johannes Gutenberg University, Mainz, Germany}
\author{G.~Li}
\affiliation{Institute for Solid State Physics, TU Vienna, Austria}
\author{K.~Held}
\affiliation{Institute for Solid State Physics, TU Vienna, Austria}
\author{N.~Bl\"umer}
\affiliation{Institute of Physics, Johannes Gutenberg University, Mainz, Germany}
\date{\today}
  \begin{abstract}
We compute the self-energy for the half-filled Hubbard model on a square lattice using lattice quantum Monte Carlo simulations and the dynamical vertex approximation.  
The self-energy is strongly momentum-dependent, but it can be parametrized via the non-interacting energy-momentum dispersion $\varepsilon_\vk$, except for pseudogap features right at the Fermi edge. That is, it can be written as $\Sigma(\varepsilon_\vk,\omega)$, with two energy-like parameters ($\varepsilon$, $\omega$) instead of three ($k_x$, $k_y$ and $\omega$). The self-energy has two rather broad and weakly dispersing high energy features and a sharp $\omega= \varepsilon_\vk$ feature at high temperatures,  which turns to  $\omega= -\varepsilon_\vk$ at low temperatures. Altogether this yields a \script{Z}- and \inverseZ-like structure, respectively, for the imaginary part of  $\Sigma(\varepsilon_\vk,\omega)$.
We attribute the change of the low-energy structure to antiferromagnetic spin fluctuations.
  \end{abstract}
  \maketitle

\section{Introduction}
\label{sec:Intro}

The calculation of strongly correlated electron systems remains
one of the biggest challenges to theoretical solid-state physics.
These correlations originate from the Coulomb repulsion between the electrons, and they strongly modify their propagation,  which is described by the single-particle Green's function.  Compared to the non-interacting  case,
the propagation  is dressed by the frequency- and momentum-dependent self-energy, $\Sigma( {\mathbf k},\omega)$. 
Thus, knowledge of the self-energy gives direct access to all single-particle properties. It allows us, in principle, also to calculate multi-particle propagators via derivatives of the self-energy with respect to appropriate source fields.\cite{mahan2013many, negele1988quantum}

Understanding the frequency and momentum structure of  $\Sigma( {\mathbf k},\omega)$ and a generic parametrization thereof is hence a  key to understanding correlated systems on the whole.  
The standard model for electrons in a solid, {\it i.e.} Landau's Fermi liquid theory,\cite{landau1957theory} corresponds to a Taylor expansion of $\Sigma( {\mathbf k},\omega)$ around the Fermi
energy $\omega=\varepsilon_F$ and around the Fermi surface ${\mathbf k} \in {\rm FS}$. 
It simplifies the self-energy to its linear terms with respect to 
  $\omega-\varepsilon_F$ and the perpendicular component of the ${\mathbf k}$ deviation from the  Fermi surface. This linearization 
allows for describing the one-particle physics at low energies of almost all metals; Mott insulators,\cite{Mott49a,Mott68a,Gebhard97a} on the other hand,  have a diverging self-energy around $\omega=\varepsilon_F$, while  non-Fermi liquids deviate from a linear dependence in  $\omega-\varepsilon_F$ despite being metallic. \cite{Stewart,Schulz} A hallmark of a strongly correlated Fermi liquid is 
also  a  kink  in  $\Sigma(\omega)$ \cite{Nekrasov,Byczuk} at the Kondo temperature  of the lattice \cite{Held13}. This kink separates a
 first linear (Fermi liquid) behavior of  $\Sigma(\omega)$ from
a second   one at larger $\omega$, where one still has quite coherent
 quasiparticle-like excitations. 

These examples already point to the importance of the $\omega$ dependence
in our present understanding of the self-energy of strongly correlated electrons.
Much less is known regarding its generic ${\mathbf k}$-dependence. This imbalance is
particularly obvious in  dynamical mean-field theory (DMFT) \cite{DMFT,DMFTreview} which 
neglects the ${\mathbf k}$ dependence altogether and considers solely $\Sigma(\omega)$. The fact that DMFT is exact in infinite dimensions \cite{Metzner1989} and
a good approximation for electronic correlations in  three dimensions, at least at elevated temperatures, justifies this imbalance in many situations. 
It is well-known, however, that the ${\mathbf k}$ dependence of $\Sigma({\mathbf k},\omega)$ is generally important in two- or one-dimensional systems.

In certain situations, the self-energy can
be separated into a frequency-dependent local and a static momentum-dependent contribution,\cite{Schaefer15}
$\Sigma({\mathbf k},\omega)= \Sigma_{loc}(\omega)+ \Sigma^{\prime}( {\mathbf k})$, which is particular appealing when these two contributions
are calculated by different methods such as DMFT and $GW$, respectively. \cite{Tomczak12,Taranto13,Tomczak14}
In general, however, as shown in different diagrammatic extensions,\cite{Rubtsov08, Rohringer2013, Li2015} the non-local part of the self-energy 
is frequency-dependent, {\it i.e.} $\Sigma(\vk,\omega) = \Sigma_{loc}(\omega) + \Sigma^{\prime}(\vk,\omega)$. 

In this paper, we show that a simpler form of the 
  ${\mathbf k}$ dependence,  
$\Sigma( {\mathbf k},\omega ) \approx \Sigma( \varepsilon_{\mathbf k},\omega),$
is possible  to a large extent. That is, $\Sigma( {\mathbf k},\omega)$ 
depends on  ${\mathbf k}$ only through the corresponding non-interacting energy $\varepsilon_{\mathbf k}$ instead of the full $d$-dimensional ${\mathbf k}$-vector. 
Such a form is obviously correct (at least) for one-dimensional systems with nearest-neighbor hopping, {\it i.e.}, monotonic $\varepsilon_{\mathbf k}$ within the reduced Brillouin zone (BZ); as well as in infinite dimensions where $\Sigma( {\mathbf k},\omega)\equiv \Sigma(\omega)$.
In two dimensions, this means that $\Sigma$ only depends on two variables ($\varepsilon$ and $\omega$) instead of three ($k_x$, $k_y$, and $\omega$).
 Even in this most 
ambitious case, our cutting-edge Blankenbecler-Sugar-Scalapino quantum Monte Carlo (BSS-QMC) simulations \cite{PhysRevD.24.2278} and dynamical vertex approximation (D$\Gamma$A) \cite{Toschi2007,Katanin2009} calculations collapse by-and-large 
onto a single $\varepsilon_\vk$-dependence in addition to the $\omega$ dependence. 
An important exception is the so-called pseudogap phase in which the
self-energy has, as a matter of course, different values
at the Fermi surface in the nodal and antinodal direction. 
In the pseudogap phase angle-resolved photoemission spectroscopy (ARPES) data show spectral weight in the nodal but not in the antinodal direction.
 But even in this phase, the single  $\varepsilon_\vk$-parametrization  is restored soon when going  away from the Fermi surface.  
Limitations also arise in the case of strongly asymmetric lattices with different hopping in the $x$- and $y$-direction.

In the following, we present the one-band Hubbard 
model and the employed methods in \refs{ModelAndMethod}. In \refss{CollapseDQMC} and \refss{CollapseDGA},
the collapse of the self-energy onto the single $\varepsilon_\vk$-dependence is shown numerically by BSS-QMC and D$\Gamma$A, respectively.
A simple parametrization of the
$\omega$- and  $\varepsilon_\vk$-dependence of the self-energy is provided in \refss{ParametrizationExplanation}.
This also allows us to gain,  in  \refss{physics}, a better understanding of the essential features of the self-energy and its  global structure  in the $(\varepsilon,\omega)$ space. 
\refss{AnisotropicCase} discusses the case of an asymmetric lattice, and \refss{Doping} examines the doped Hubbard model. Finally,  \refs{ConclusionAndOutlook} provides a summary and an outlook.

\section{Model and Methods}
\label{sec:ModelAndMethod}
In this paper, we consider the single-band Hubbard model on a square lattice to gain insight into the $\varepsilon_\vk$ dependence of the self-energy and for the parametrization thereof. A similar analysis to what will be shown in the rest of this work can be  applied to other correlated models straightforwardly.
The single-band Hubbard model consists of two competing terms, {\it i.e.} the kinetic energy and the interaction, which describe the two most fundamental processes as well as the most essential energy scales of this model. 
Due to the competition of these two terms, more energy scales can emerge, such as the spin fluctuations appearing at low temperatures characterized by the energy scale $J\sim 4t^{2}/U$.  
The self-energy, as a measure of correlations, is certainly expected to contain these energy scales. 
Thus, a better understanding of the self-energy structure can help to understand them, as well as their competitions.
In addition to the trivial energy scales characterized by $t$ and $U$, one aim of this work is to identify the emerging energy scale, such as $J$, from the self-energy of this model.  

The single-band Hubbard model is given by
\begin{equation}\label{Hamiltonian}
H = -t\sum_{\langle i,j\rangle,\sigma}(c_{i\sigma}^{\dagger}c^{}_{j\sigma}+h.c.) -\mu\sum_{i}n_{i}+ U\sum_{i}n_{i\uparrow}n_{i\downarrow}\;,
\end{equation}
where $\langle i, j\rangle$ restricts the single-particle hopping to  nearest-neighbor sites $i$ and $j$; $c_{i\sigma}^{\dagger}$  and $c^{}_{j\sigma}$ denote the corresponding creation and annihilation operators with spin $\sigma$, respectively.
In the case that site $i$ is occupied by a spin-up electron, an energy barrier of Coulomb repulsion $U$ has to be overcome for a spin-down electron to hop onto it and vice versa.

Despite the simplicity of this model, solving it imposes a great challenge to theory, especially in two dimensions.
In this work, we will consider the Hubbard model in \refqq{Hamiltonian} on a square lattice at half-filling and solve it with two complementary approaches, {\it i.e.} BSS-QMC~\cite{PhysRevD.24.2278} and the D$\Gamma$A~\cite{Toschi2007}. Let us note that cluster extensions of DMFT can also incorporate $\mathbf{k}$-dependences of the self-energy, and they have been used intensively for the two-dimensional (2D) Hubbard model.\cite{PhysRevB.58.R7475, PhysRevLett.87.186401, PhysRevB.62.R9283, PhysRevB.61.12739, PhysRevB.64.195130, RevModPhys.77.1027} For further methods and comparisons, see Ref.~\onlinecite{PhysRevB.83.075122, PhysRevX.5.041041}.
The BSS-QMC is numerically exact for a given cluster size used in the calculations. Thus, all short-range correlations inside the cluster can be faithfully described. 
In contrast, in D$\Gamma$A both short- and long-range correlations are included, as it works in the thermodynamic limit of the problem. However,  D$\Gamma$A relies on the approximation that the irreducible vertex is local.
Thus, BSS-QMC and D$\Gamma$A are complementary to each other with respect to correlation lengths, which implies that an agreement between both methods rules out significant finite-size effects.\cite{Schaefer_Fate2015}

In the following, we briefly review the basic idea of these two approaches. For readers who are familiar already with their methodologies the rest of this section can be safely skipped.

\subsection{BSS-QMC}
The BSS-QMC algorithm is based on two important transformations, the Suzuki-Trotter decomposition~\cite{Suzuki01111976, Trotter} and the discrete Hubbard-Stratonovich transformation.\cite{stratonovich, PhysRevLett.3.77, PhysRevB.28.4059}
With the aid of the former, the kinetic ${\cal K}$ ($t$) and interaction ${\cal V}$ ($U$) terms of \refqq{Hamiltonian} can be separated  in the partition function by
\begin{eqnarray}
e^{-\beta({\cal K} + {\cal V})}
&=& \left[e^{-\Delta\tau{\cal K}}e^{-\Delta\tau{\cal V}}\right]^{L} + {\cal O}\left[(\Delta\tau)^{2}\beta\right]\:,
\end{eqnarray}
where $\Delta\tau=\beta/L$ is the discretized imaginary time. The second term is neglected in BSS-QMC, which represents the only error (besides the  statistical error of the stochastic sampling) of this method. 
This error can however be addressed by extrapolating $\Delta\tau\rightarrow0$, which is done in practice via a multigrid procedure.\cite{PhysRevA.80.051602, PhysRevE.87.053305}

The kinetic energy term, ${\cal K}$, is quadratic in terms of fermionic operators, and its exponential can be easily calculated. In contrast,
the interaction term ${\cal V}$ is quartic. One can transform it to a quadratic dependence on the fermionic operator through the Hubbard-Stratonovich transformation at the cost of introducing a discrete Ising variable $s=\pm1$ via
\begin{equation}
e^{-U\Delta\tau n_{i\uparrow}n_{i\downarrow}} = \frac{1}{2}\sum_{s=\pm 1}\prod_{\sigma=\uparrow,\downarrow}e^{-c_{i\sigma}^{\dagger}\left(\sigma s\lambda+\frac{U\Delta\tau}{2}\right)c^{}_{i\sigma}}\;,
\end{equation}
where $\lambda=\mbox{acosh}\left(|U|\Delta\tau/2\right)$\;.

With both the kinetic and the potential terms in their quadratic forms, the partition function can be simply written as a product of two determinant matrices
\begin{equation}\label{partition_function}
{\cal Z}=\left[\frac{e^{-U\Delta\tau/4}}{2}\right]^{NL}\mbox{Tr}_{\{s_{i}\}}\det\left[M_{\uparrow}(s)\right]\det\left[M_{\downarrow}(s)\right]\;,
\end{equation}
where $s=\left(s_{1}, s_{2}, \cdots, s_{L}\right)$ and
\begin{equation}
M_{\sigma}(s)=I+B_{L,\sigma}(s_{L})B_{L-1,\sigma}(s_{L-1})\cdots B_{1,\sigma}(s_{1})\;.
\end{equation}
Here, the matrix $B_{l,\sigma}(s_{l})$ is defined from the quadratic forms of the kinetic and potential energies $B_{l,\sigma}(s_{l})=e^{-\Delta\tau \tilde{{\cal K}}}e^{-\Delta\tau \left(\sigma s_{l}\lambda/\Delta\tau+ U/2\right)}$ where ${\tilde{\cal K}}$ denotes the kinetic energy matrix.
With the partition function in \refqq{partition_function}, other thermodynamic observables, such as the single-particle Green's function, can be readily obtained. It requires a Monte Carlo sampling over the Ising-field $s$. For more details of the BSS-QMC, we refer the reader to the review works~\onlinecite{assaad2008world, chang2015recent}.
In this work, we will take the self-energy calculated from BSS-QMC as a reference and analyze its momentum-energy structure to gain insight and to derive   a parametrization of it.

\subsection{\protect\boldmath D$\Gamma$A}
\label{subsec:dgamma}
 D$\Gamma$A takes a different strategy. It is one of several recent  diagrammatic extensions of DMFT\cite{Kusunose06,Toschi2007,Katanin2009,Rubtsov08,Hafermann2008,Rohringer2013,Taranto2014,Li2015,Ayral2015,Kitatani2015} that start from a local two-particle vertex and generate local and non-local correlations from it.  D$\Gamma$A assumes the irreducible vertex to be local, starting either from  the irreducible vertex in a given, e.g., particle-hole channel \cite{Toschi2007,Katanin2009} or the fully irreducible vertex \cite{Valli2015,Li2015b}. In the former case, local and non-local correlations are generated by the Bethe-Salpeter equation  \cite{Toschi2007,Katanin2009}; in the latter case this is done by the parquet equation \cite{Valli2015,Li2015b}. Here, we follow the implementation \cite{Katanin2009} that first calculates the local particle-hole-irreducible vertex $\Gamma_{s(c),{\bf ir}} (\nu,\nu',\omega)$ in the spin (charge) channel from  a converged DMFT solution.
From this local irreducible vertex, the full vertex (or the  susceptibility) is calculated through the Bethe-Salpeter ladder, including a Moriyaesque $\lambda$-correction to mimic self-consistency effects. The full vertex also includes the corresponding  transversal particle-hole ladder diagrams, which do not need to be calculated explicitly since they follow by crossing symmetry from the particle-hole ladder diagrams.
 The full vertex, in turn, allows us to calculate the self-energy through the exact Heisenberg equation of motion which is also known as Schwinger-Dyson equation.
 For a pedagogical introduction we refer the reader to Ref.\ \onlinecite{Held2011}. Here, we only recall the most important ladder D$\Gamma$A equations, cf. Ref. \onlinecite{Toschi2007,Katanin2009,Rohringer2011}.

The ${\mathbf q}$-dependent susceptibility $\chi _{\mathbf{q}\omega }$ is calculated from the Bethe-Salpeter equation  as 
\begin{equation} 
\chi _{\mathbf{q}\omega }^{s(c)}=\left[\left(\phi _{\mathbf{q}\omega 
}^{s(c)}\right)^{-1}\mp U+\lambda _{s(c)}\right]^{-1},
\end{equation}%
where, $\lambda _{s(c)}$ is the   Moriyaesque $\lambda$-correction in the spin (charge) channel (it is actually relevant only in the spin-channel here);  and
$\phi _{\mathbf{q}\omega }^{s(c)} =\sum\limits_{\nu \nu ^{\prime }}\Phi_{s(c),\mathbf{q}}^{\nu \nu ^{\prime }\omega }$  is obtained from the local irreducible vertex $\Gamma$\footnote{For the properties of the local vertices see Ref.~\onlinecite{PhysRevB.86.125114, PhysRevLett.110.246405}} 
through the Bethe-Salpeter equation as
\begin{eqnarray} 
\Phi _{s(c),\mathbf{q}}^{\nu \nu ^{\prime }\omega } &=&\left[\left(\chi _{0,\mathbf{q}\omega 
}^{\nu ^{\prime }}\right)^{-1}\delta _{\nu \nu ^{^{\prime }}}-\Gamma _{s(c),{\bf ir}}^{\nu \nu ^{\prime }\omega }\pm U\right]^{-1}.
\end{eqnarray}%

Here, 
 $\chi _{0,\mathbf{q}\omega }^{\nu ^{\prime }}=-T\sum_{\mathbf{k}}G_{\mathbf{k} 
,\nu ^{\prime }}G_{\mathbf{k}+\mathbf{q},\nu ^{\prime }+\omega }$ is the  
particle-hole  bubble susceptibility; 
 $G_{\mathbf{k},\nu }=\left[i\nu -\varepsilon _{\mathbf{k}}+\mu -\Sigma _{\text{loc}}(\nu )\right]^{-1}$ the ($\mathbf{k}$-dependent) DMFT Green's function;
$\Sigma _{\text{loc}}(\nu )$ the DMFT self-energy; and $T=1/\beta$ the temperature.

The  Schwinger-Dyson equation for the ${\bf k}$- and $\omega$-dependent self-energy 
reads\cite{Katanin2009}
\begin{eqnarray} 
\Sigma(\mathbf{k},\nu) &=&\frac{1}{2}{Un}+\frac{1}{2}TU\sum\limits_{%
\omega ,\mathbf{q}}\Big[ 3\gamma _{s,\mathbf{q}}^{\nu \omega }-\gamma _{c,%
\mathbf{q}}^{\nu \omega }-2   \nonumber \\ 
&&+3U\gamma _{s,\mathbf{q}}^{\nu \omega }\chi _{\mathbf{q}\omega 
}^{s}+U\gamma _{c,\mathbf{q}}^{\nu \omega }\chi _{\mathbf{q}\omega }^{c}  
\nonumber \\ 
&&+\sum\limits_{\nu ^{\prime }} \chi _{0,\mathbf{q}\omega }^{\nu 
^{\prime }}(\Gamma _{c,\text{loc}}^{\nu \nu ^{\prime }\omega }-\Gamma _{s,%
\text{loc}}^{\nu \nu ^{\prime }\omega })\Big] G_{\mathbf{k+q},\nu +\omega },
\label{SE0} 
\end{eqnarray}%
where  
$\gamma _{s(c),\mathbf{q}}^{\nu \omega }=(\chi _{0,\mathbf{q}\omega }^{\nu
})^{-1}\sum\limits_{\nu ^{\prime }}\Phi _{s(c),\mathbf{q}}^{\nu \nu ^{\prime 
  }\omega}$, 
$n$ is the particle number. The prefactor 
$3$ and $1$ for the s(pin) and c(harge) $\gamma _{s(c),\mathbf{q}}$
already includes the particle-hole-transversal ladder which for the one-band Hubbard model only enters for the spin channel. The local 
$\Gamma _{s,  \text{loc}}^{\nu \nu ^{\prime }\omega }$ terms subtract those diagrams contained in both, particle-hole and particle-hole transversal channel. This ladder D$\Gamma$A neglects the particle-particle channel, which is of less relevance in the spin-fluctuation dominated range of the phase diagram, at least close to the Fermi energy.\cite{Valli2015}

\section{Results}
\label{sec:Results}

By carefully examining the structure of the self-energy calculated by the BSS-QMC and the D$\Gamma$A, we want to show  in the following that a simplified $\vk$-dependence of the self-energy~-- via the non-interacting dispersion $\varepsilon_{\vk}$~-- can be achieved. 
To see the advantage of such a parametrization, let us first examine the self-energy in the full momentum-frequency space, {\it i.e.} as a function of $k_{x}$, $k_{y}$ and $\omega_{n}$. 
Initially, we restrict ourselves to the case of isotropic hopping on a square lattice and half-filling; for generalizations, see \refss{AnisotropicCase} and \refss{Doping}, respectively. Results for the intermediate coupling $U=4t$ are collected in \reff{Sigma_along_k}.
In this coupling regime and at a temperature of $\beta t = 5.6$, the system is in the regime where the pseudogap opens. 
At lower temperatures, the paramagnetic phase becomes insulating\cite{Schaefer_Fate2015} and eventually also antiferromagnetic at $T=0$.

\begin{figure}[t]
  \includegraphics[width=\columnwidth]{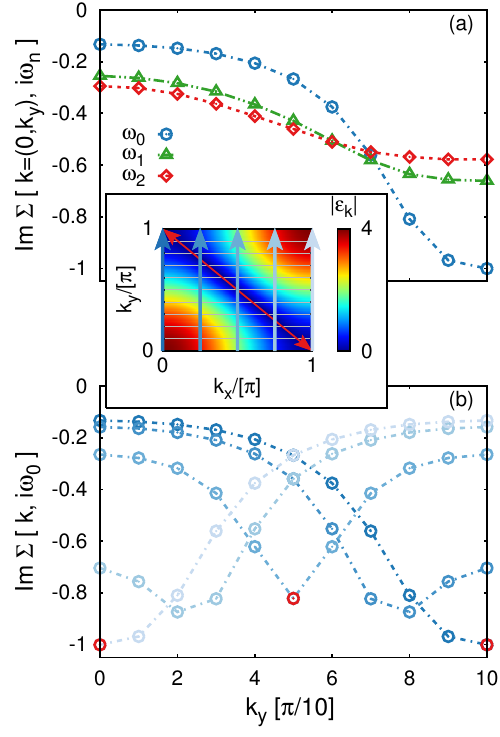}
  \caption{(Color online):
 Imaginary part of the self-energy $\Sigma(\vk,\wn)$ from BSS-QMC for $U=4t, $ and $\beta t =5.6$ at (a)  the first three Matsubara frequencies and $k_{x}=0$; (b) at the first Matsubara frequency along the (brightness-coded) five momentum paths shown in the inset. The red points in (b) correspond to the nodal and antinodal point, which are emphasized alike in the inset by the red diagonal arrow.}
  \label{fig:Sigma_along_k}
\end{figure}

The upper panel, 
 \reffa{Sigma_along_k}, shows the imaginary part of the self-energy at $\vk=(0,k_{y})$ for the first three Matsubara frequencies as a function of $k_y$.
 The variations along this high-symmetry cut through the BZ are seen to be quite significant, by a factor of about 10 at $\omega_0$ (circles) and still by a factor of about two at $\omega_2$ (diamonds). Evidently, DMFT would be completely inadequate in this respect. Only at large frequencies does the self-energy become asymptotically momentum-independent: $\Sigma(\vk,i\omega_n)\stackrel{\omega_n\to\infty}{\longrightarrow}U^2 /(4 i \omega_n)$.  

As seen in \reffb{Sigma_along_k}, the self-energy at $\Sigma(\vk,\omega_{0})$ varies strongly also along the other momentum paths indicated in the inset of \reff{Sigma_along_k}, without an obvious structure (except for the evident mirror symmetry line $k_x=k_y$). 
This dependence is not particularly smooth, on the scale of our momentum grid. This indicates that approximations of the self-energy by piecewise constant patches, as usually employed in the dynamical cluster approximation (DCA)~\cite{RevModPhys.77.1027} (on much coarser grids), may be problematic for small cluster sizes.
Instead, accurate approximation schemes would have to incorporate insight in the momentum structure of $\Sigma$ 
[or use expansions 
of the self-energy that are not stepwise constant, such as the cumulant expansion (see Ref.~\onlinecite{PhysRevB.74.125110, Stanescu20061682, PhysRevB.85.035102, PhysRevB.88.115101, PhysRevLett.102.056404, PhysRevLett.116.057003})].


\subsection{\protect\boldmath Collapse of $\vk$-dependence to a $\varepsilon_{\vk}$-dependence}
\label{subsec:CollapseDQMC}

\begin{figure}[t]%
  \includegraphics[width=\columnwidth]{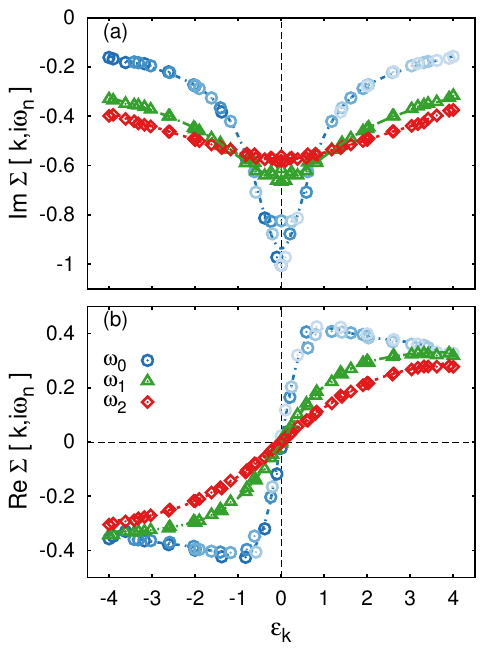}%
  \caption{(Color online): Imaginary (a) and real (b) part of the self-energy $\Sigma(\vk,\wn)$ vs. the non-interacting dispersion $\varepsilon_{\vk}$ from BSS-QMC at $U = 4t$ and $\beta t= 5.6$. Different $(k_x,k_y)$ points with the same $\varepsilon_{\vk}$  collapse onto a single curve.}
\label{fig:Sigma_eps_iwn}%
\end{figure}%


\reffa{Sigma_eps_iwn} shows Im~$\Sigma(\vk,i\omega_n)$ at the first three Matsubara frequencies plotted versus the non-interacting single-particle energy ({\it i.e.}, band dispersion)  $\varepsilon_\vk$. 
Specifically, the circles in \reffa{Sigma_eps_iwn} represent all the data of \reffb{Sigma_along_k} (corresponding to the lowest Matsubara frequency $\omega_0$).
Not surprisingly, this data set is peaked for $\vk$ at the Fermi edge, $\varepsilon_{\vk}=0$; this is also true at the higher frequencies $\omega_1$ (triangles) and $\omega_2$ (diamonds). However, it is remarkable that each of these data sets collapse on a single line with high accuracy, with the exception of only a very narrow region around $\varepsilon_\vk=0$. Global collapses are also seen in the corresponding real parts, shown in \reffb{Sigma_eps_iwn}; here no low-$\varepsilon_\vk$ deviations can be seen due to the linearity of Re~$\Sigma$ at low $\varepsilon_\vk$.

The significant momentum dependence of Im~$\Sigma(\vk,\omega)$ at $\varepsilon_\vk=0$, on the other hand, is nothing short of the pseudogap physics exposed in cluster extensions of DMFT,\cite{RevModPhys.77.1027, PhysRevLett.95.106402, PhysRevLett.111.107001} recent BSS-QMC\cite{Rost2012} and D$\Gamma$A studies\cite{Schaefer_Fate2015}: the self-energy takes different values at the Fermi surface along the nodal and antinodal directions, with variations of about $20\%$. In this respect, the nodal and antinodal points are highlighted in \reffb{Sigma_along_k} as well as in the inset thereof. We learn from \reffa{Sigma_eps_iwn} that this physics is, however, narrowly confined to momentum space around the Fermi surface. Note that nodal/antinodal variations of the gap decay quickly both towards higher and lower temperatures.\cite{Rost2012} In this sense, the parameter choice of \reff{Sigma_eps_iwn} ($U=4t, \beta t =5.6$)  may be considered a worst case  for parametrizing $\Sigma$ via $\varepsilon_\vk$.
%
%
%
\begin{figure}[t]%
  \includegraphics[width=\columnwidth]{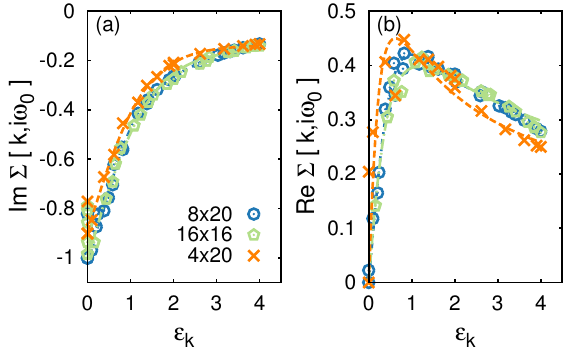}%
  \caption{(Color online): Imaginary (a) and real (b) part of the self-energy $\Sigma(\vk,i\omega_0)$ from BBS-QMC at $U=4t$ and $\beta t=5.6$. The different data points correspond to different lattice sizes and geometries.}%
  \label{fig:Sigma_eps_iwn_inset}%
\end{figure}%

The collapse of the self-energy  $\Sigma(\vk,i\omega_0)$ onto a single $\varepsilon_{\vk}$-dependent  $\Sigma(\varepsilon_{\vk},i\omega_0)$ remains unchanged when changing the cluster size in the BSS-QMC calculations. \reff{Sigma_eps_iwn_inset} compares results obtained for different system sizes and geometries. In this respect, the self-energy at $U=4t$ and $\beta t=5.6$ is shown for two lattice systems with rectangular shape, $4\times20$ and $8\times20$, as well as for a system with a regular square shape, having $16\times16$ sites. We find that (i) the collapse onto a single curve (versus $\varepsilon_\vk\not=0$) is better for larger systems and (ii) that, overall, the convergence seems to be quite rapid,
which justifies in a qualitative way that we  skipped the interpolation to an infinite system. 
The conclusion that our analysis is relevant in the thermodynamic limit will be further verified in Sec.~\ref{subsec:CollapseDGA} by comparing BSS-QMC data, as well as the self-energy parametrization discussed in the next section, with the results obtained from $\DGA$. 

Very importantly, the collapse of data points with respect to $\varepsilon_{\vk}$ is not restricted to certain interaction strengths.
The self-energy in different phases (bad-metallic towards insulating) characterized by different values of $U$ is shown in \reff{Sigma_eps_iwn_U}. 
Compared to the case of $U=4t$ where the phase transition approximately occurs (see Ref.~\onlinecite{Rost2012}), the $\vk$ variations scale with a factor of 15 for $U=8t$; and a factor of 0.1 for $U=2t$.
Despite the stronger  $\varepsilon_{\vk}$-dependence at the larger-$U$ regime, all data still collapse onto a single curve. That is, our parametrization discussed in the next section can be equally applied to both, the weak- and strong-coupling regime. It is not  perturbative.

\subsection{\protect\boldmath Parametrization of $\Sigma$}
\label{subsec:ParametrizationExplanation}

\begin{figure}[t] 
	\includegraphics[width=\columnwidth]{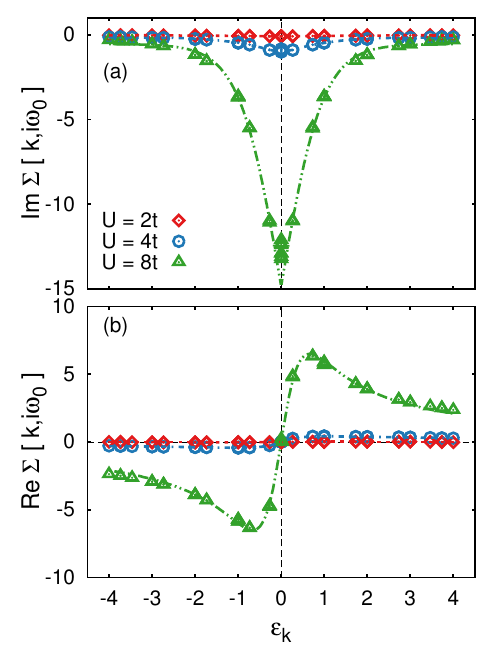}
	\caption{(Color online): Imaginary (a) and real (b) part of the self-energy $\Sigma(\vk,i\omega_0)$ from BSS-QMC at $\beta t= 5.6$ for different $U$-values.}
	\label{fig:Sigma_eps_iwn_U}
\end{figure}

So far, we have discussed self-energies on the imaginary frequency axis, following a common practice especially within the QMC community. While such data have the advantages of direct accessibility from (imaginary-time) QMC data and easier comparisons with literature data, real-frequency results are obviously more physically relevant and also more interesting. Such data, obtained via maximum entropy analytic continuation~\cite{Jarrell1996133, bergeron2015algorithms} $i \omega_{n}\rightarrow\omega$ on the level of the self-energy, are shown in \reffa{Sigma_GF_B056} as a function of $\varepsilon_{\vk}$ and $\omega$ at $\beta t=5.6$ and $U=4t$; the corresponding Green's function, obtained via the Dyson equation, can be seen in \reffb{Sigma_GF_B056}. Note that these data (Im~$\Sigma$, Im~$G$) are, up to factors $-\pi$, spectral functions that also fully determine the corresponding real parts via Kramers-Kronig relations.

\reffa{Sigma_GF_B020} and (b) show the same quantities but at a higher temperature $\beta t=2$.

\begin{figure}[t]
	\includegraphics[width=\columnwidth]{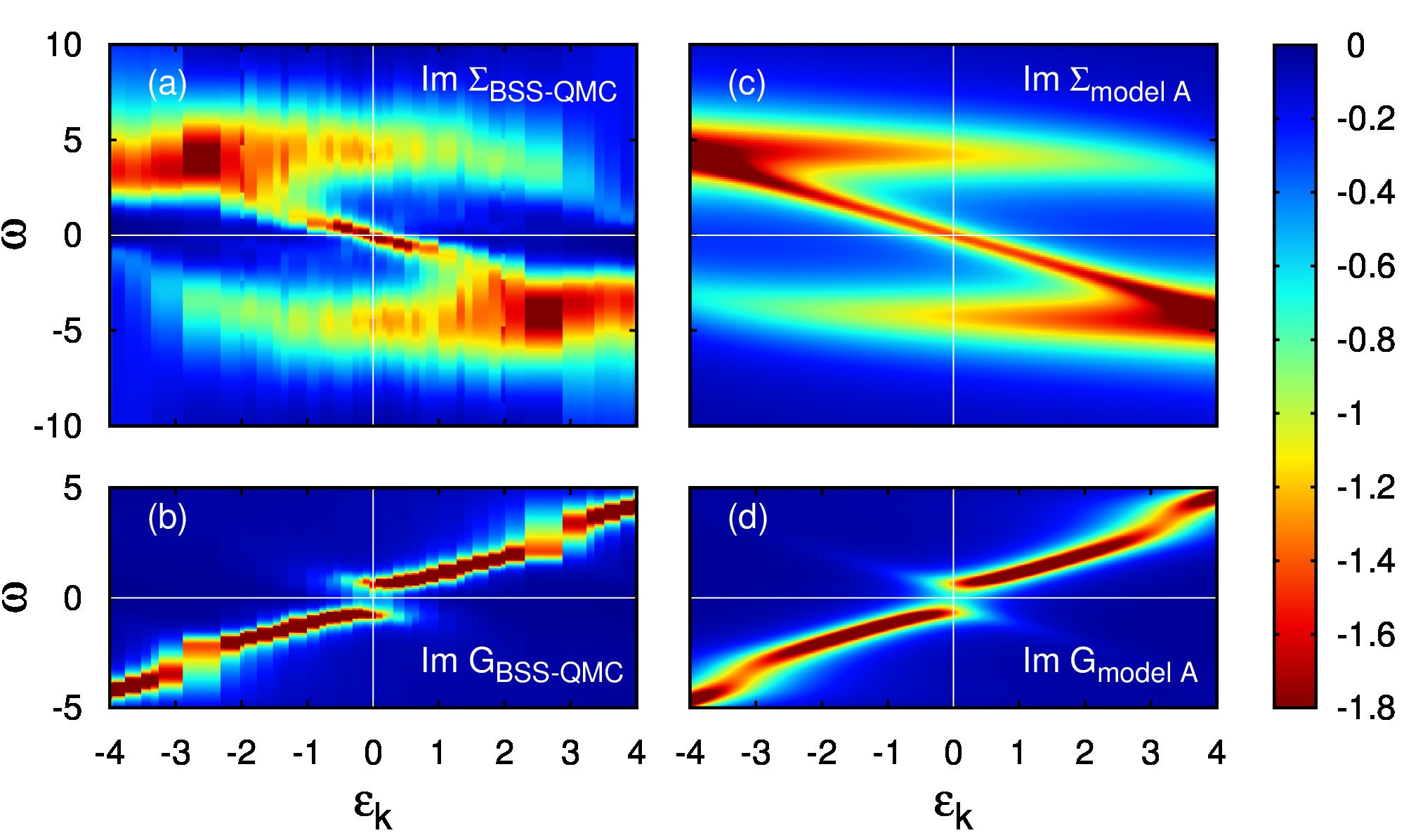}
	\caption{(Color online): Imaginary part of the self-energy $\Sigma(\varepsilon,\omega)$ and of the Green's function $G(\varepsilon,\omega)$ at $U = 4t$ and $\beta t= 5.6$. (a) and (b) contain the BSS-QMC data. (c) and (d) represent continuous parametrizations, denoted in \refq{modelFct} and \reft{ModelParameters}.}
	\label{fig:Sigma_GF_B056}
\end{figure}

\begin{figure}[t]
\includegraphics[width=\columnwidth]{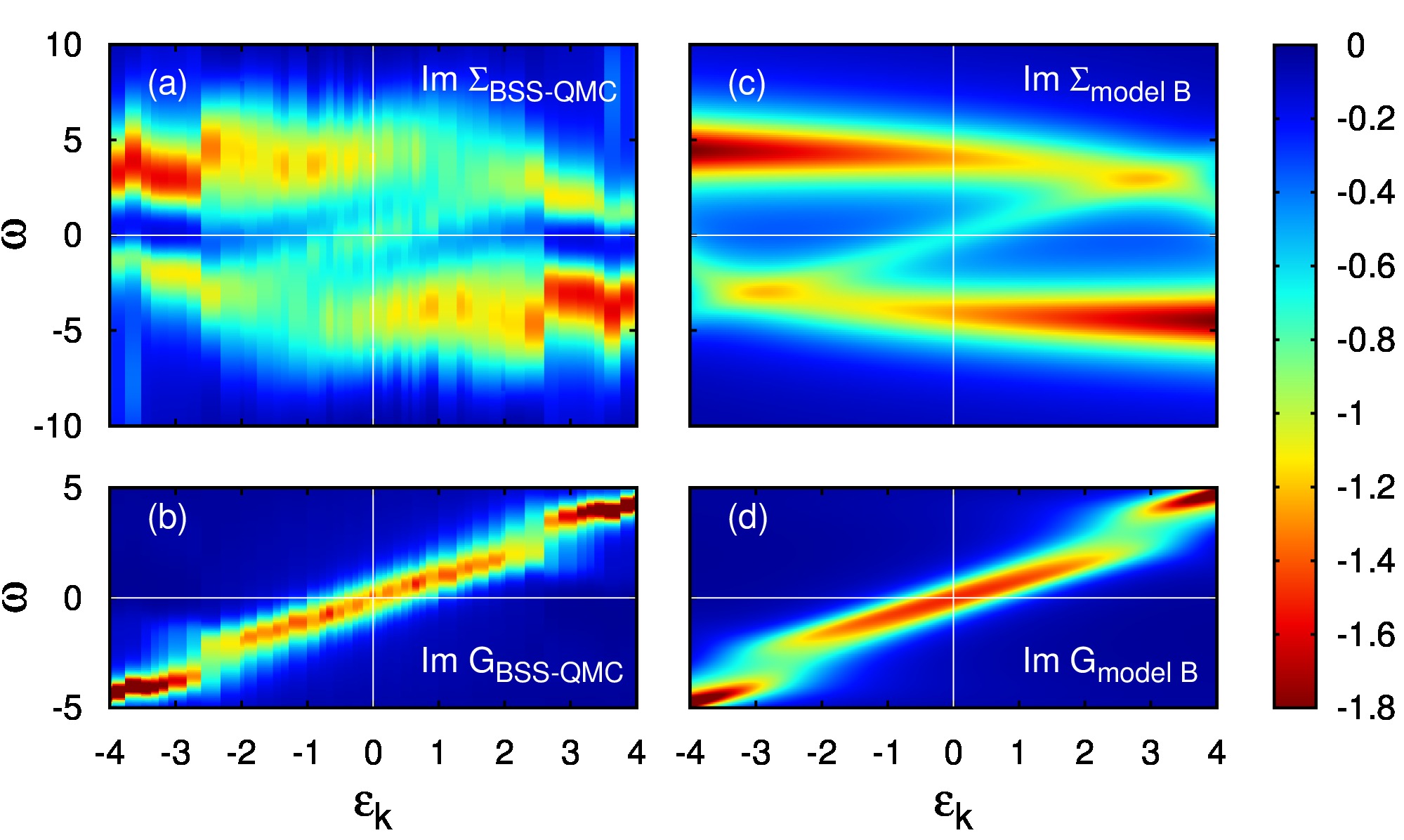}
\caption{(Color online):  Same as  \reff{Sigma_GF_B056} but  at a higher temperature, $\beta t=2$.}
\label{fig:Sigma_GF_B020}
\end{figure}

\begin{figure}[b]
	\includegraphics[width=\columnwidth]{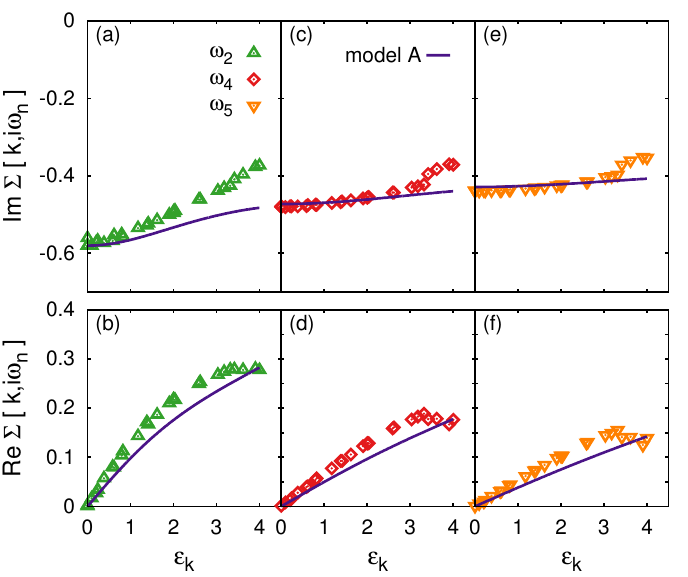}
	\caption{(Color online): The self-energy at three different (higher) Matsubara frequencies at $U = 4t$ and $\beta t= 5.6$. 
Together with the asymptotic behavior of the Matsubara self-energy, these BSS-QMC data are used to fix the model parameters in \refqq{eq:modelFct}. 
The fitted parameters can be found in \reft{ModelParameters}.}
	\label{fig:fitOfParameters}
\end{figure}

Let us now discuss the structures seen in \reffa{Sigma_GF_B056} and \reffa{Sigma_GF_B020}. Both share a common feature, namely  broad bands at high frequencies (both positive and negative), which are nearly dispersionless, {\it i.e.} with maxima fixed at $|\omega|\approx 4$. 
However, at a given $\varepsilon_\vk$ these two dispersionless branches do not have the same weight. That is, with increasing $\varepsilon_\vk$ (for $\varepsilon_\vk>0$) spectral weight from the upper band ($\omega\gtrsim2$) is shifted towards the lower band ($\omega\lesssim -2$); and vice versa for $\varepsilon_\vk<0$.
In addition to this high-energy structure, a strong low-energy feature with negative slope is seen at the lower temperature, in the pseudogap phase in \reffa{Sigma_GF_B056} (precisely at $\omega=-\varepsilon_\vk$). 
 Overall, this implies a \inverseZ shaped spectral distribution of the self-energy.\cite{Sakairef} Its low-energy part
splits the (Green's function) spectral density at $\omega\approx 0$, $\varepsilon_\vk\approx 0$, {\it i.e.}, introduces the pseudogap [see \reffb{Sigma_GF_B056}].

Consequently, it is clear that the overall structure must be different at higher energies, above the pseudogap phase. However, it is surprising that the diagonal with negative unit slope, observed before, is completely absent (instead of only being weakened) in \reffa{Sigma_GF_B020} and replaced by another diagonal with positive unit slope, {\it i.e.}, with maxima at $\omega=\varepsilon_\vk$. As seen in \reffb{Sigma_GF_B020}, this leads to a (Green's function) spectral density that is only broadened in a wide frequency range, but gap-less, {\it i.e.} not split at $\omega\approx 0$.


To faithfully model the structure of the self-energy at both low and high temperatures, we consider the following parametrization, 
\begin{equation}
  \Sigma(\varepsilon,\omega) = \frac{m_1}{\omega+s\varepsilon +id_1/2} + \sum_{\alpha=\pm}\frac{m_2f_{\alpha}(\varepsilon)}{\omega+h_{\alpha}(\varepsilon)+id_2/2}\;,
  \label{eq:modelFct}
\end{equation}
which is obtained by decomposing the self-energy  \reffa{Sigma_GF_B056} [\reffa{Sigma_GF_B020}] into three key features (components): the two horizontal stripes (antisymmetric in $\varepsilon$) and one sharp diagonal stripe with $s=+1$ [$s=-1$]. 
Each component has a density profile, which is represented by a Lorentzian function with weight $m_{1(2)}$ and width $d_{1(2)}$.
For the horizontal stripes, the functions $f_{\alpha}$ and $h_{\alpha}$ describe the $\varepsilon$-dependent weight and the degree of curvature, which are taken as $f_{\pm}(\varepsilon)=1 \pm b\,\varepsilon$ and $h_{\pm}(\varepsilon)=\pm5\frac{c\pm\varepsilon}{c\pm\varepsilon+1}$.

Please note the plus and minus sign ($s=\pm1$) in front of $\varepsilon$ in the first term of \refq{modelFct}. Depending on the temperature, the physics is quite different as discussed in \refss{physics} below. This is reflected in the two different signs.
At low temperatures (model A\cite{Sakairef})
we have the plus sign ($s=+1$), and at high  temperature (model B) we have the minus sign ($s=-1$) for the first term of \refq{modelFct}.
To fix the parameters, we first require that the model function in \refq{modelFct} behaves asymptotically as $U^{2}/(4\omega)$ for $\omega\rightarrow\infty$,  which reduces the independent parameters of the model function by one  ($m_1+2m_2=4$). 
The rest of the parameters are then determined by fitting the Matsubara self-energy with a least-squares approach, as shown in \reff{fitOfParameters}.
In \reft{ModelParameters}, we list the different parameters of $m_{1}, d_{1}, d_{2}, b$ and $c$,  for the low- and high- temperature phases of the Hamiltonian in \refqq{Hamiltonian}.

\begin{table}[t]
\squeezetable
\begin{ruledtabular}
\begin{tabular}{ccccccc}
& $m_{1}$ & $d_{1}$ & $d_{2}$ & $b$ & $c$ & $s$\cr
\hline
model A & 0.6 & 1.0 & 3.0 & 0.12 & 5.8 & $+1$\cr
model B & 0.4 & 2.1 & 3.0 & 0.12 & 4.5 & $-1$\cr
\end{tabular}
\caption{Different choices of parameters for the two models derived from \refq{modelFct}. A and B correspond to the two best models for the self-energy at low ($\beta t=5.6$) and high temperatures ($\beta t=2$), corresponding to  \reff{Sigma_GF_B056} and \reff{Sigma_GF_B020}, respectively. }
	\label{tab:ModelParameters}
\end{ruledtabular}
\end{table}

Despite the simple form of \refq{modelFct}, the essential structure of the self-energy and its temperature evolution can be nicely reproduced by this parametrization. 
In \reffc{Sigma_GF_B056} and (d) the self-energy and the corresponding Green's function calculated from \refq{modelFct} are shown and compared to the numerically exact solution from BSS-QMC on a finite $\vk$ grid. 
As we can see, model A nicely reproduces the three major structures of the self-energy, including the two horizontal stripes at high energy and the linear dependence of $\varepsilon_\vk$ at low energies. 
As a result, the Green's function in model A also nicely reproduces that of the BSS-QMC shown in \reffb{Sigma_GF_B056}.   

At $\beta t=2$, we adopt the parameter set indicated as model B in \reft{ModelParameters}. The comparison of model B with the BSS-QMC results is shown in \reffa{Sigma_GF_B020}-(d). 
At this higher temperature, as clearly seen from the BSS-QMC results, the horizontal stripes at high energy remain, while the low-energy linear dependence on $\varepsilon_\vk$ completely changes its sign as compared to \reffa{Sigma_GF_B056}, which applies a strong constraint on our model function, since a correct parametrization should also faithfully reproduce the sign change on the $\varepsilon_\vk$ dependence of the self-energy at the low-energy regime. 
From \reff{Sigma_GF_B020}, we see that model B nicely generates the correct $\varepsilon_\vk$-dependence, as well as the two horizontal stripes. 

\subsection{\protect\boldmath Physics associated to the parametrization of $\Sigma$}
\label{subsec:physics}

In the following, we want to show that the observed structure with weakly temperature-dependent horizontal stripes and the strongly temperature-dependent linear low-energy features
 are natural consequences of the essential 
particle-hole excitations and the magnetic correlations of the Hubbard model on the square lattice. 
Correctly reproducing those two physical processes in our self-energy model function is a strong validation of this parametrization.  
Our model function, can thus be used to describe the low-energy excitations in both the charge and the spin sectors of this model.

\begin{figure}[t]
\centering
\includegraphics[width=\linewidth]{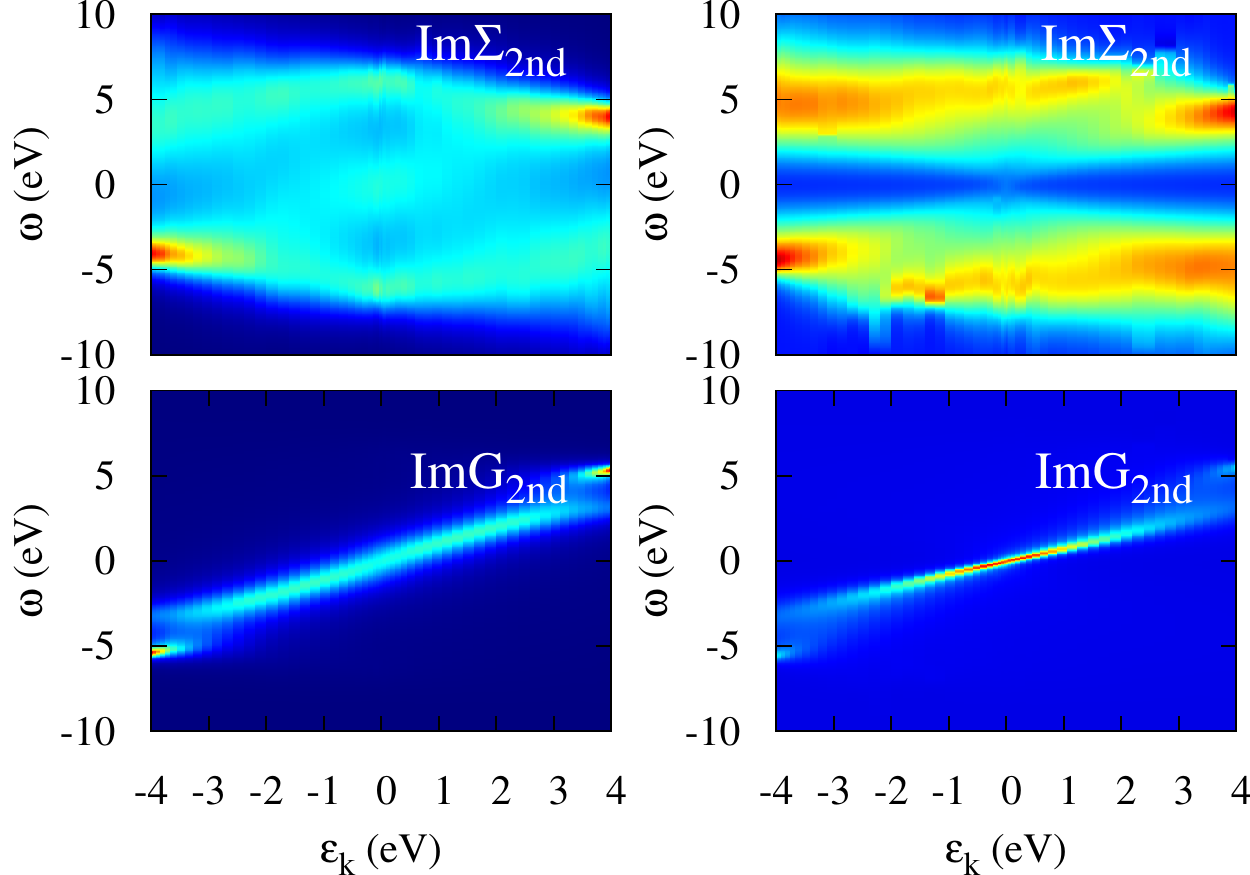}
\caption{(Color online): Self-energy and the Green's function calculated from second-order perturbation theory at $\beta t = 1$  (left) and  $\beta t = 5.6$ (right)  for $U=4t$. }
\label{fig:Sigma_2nd}
\end{figure}

We start from second-order perturbation theory of the self-energy, which effectively describes the motion of electrons in the background of particle-hole excitations, 
\begin{eqnarray}
\Sigma(\vk,\nu_n) &=& -\frac{U^2}{\beta^2N^2}\sum_{k',\nu'_n,q,\omega_n} G_{\mathbf{q}-\mathbf{k},\nu_n-\omega_n}\nonumber\\
&&\cdot G_{\mathbf{q}-\mathbf{k'},\omega-\nu'_n}G_{\mathbf{k'},\nu'_n} \:.
\end{eqnarray}
For the analytic continuation we utilize the Pad\'{e} approximation.\cite{Vidberg}
\reff{Sigma_2nd} shows the corresponding self-energy and the Green's function at two different temperatures  $\beta t=1$ (left) and $\beta t = 5.6$ (right). 
At both low and high temperatures, the self-energy from the second-order perturbation theory displays the two horizontal stripes at high energies. 
At high temperature, the same $\omega = \varepsilon_\vk$ stripe  as  in \reffa{Sigma_GF_B020} shows up. 
We thus conclude that the appearance of the horizontal stripe is due to the particle-hole excitations, which exists at both high and low temperatures. 

In the low-energy regime, the linear dependence of the self-energy on $\varepsilon_\vk$ disappears at low temperature, e.g., it can hardly be seen in \reff{Sigma_2nd} (right). 
But it is not replaced by a negative linear dependence of the self-energy on $\varepsilon_\vk$ as observed in \reffa{Sigma_GF_B056}.
This clearly tells us that the negative linear dependence in \reffa{Sigma_GF_B056} is {\it not} due to particle-hole excitations. 
We find that it is, instead, an indication of the low-temperature spin-density-wave (SDW) of the 2D Hubbard model in the self-energy function.  
To see this, we consider a mean-field description of the Hubbard model in \refqq{Hamiltonian} in the presence of a SDW.\footnote{For a review on phenomenological theories of the pseudogap in terms of such mean-field descriptions see Ref.~\onlinecite{Rice2012}}
The Fermi surface of the half-filled Hubbard model on the square lattice is nesting which favors the formation of a SDW with magnetic wave vector $\mathbf{Q}=(\pi, \pi)$.
The corresponding magnetic Brillouin zone (MBZ) is, then, only half of the original BZ, so that the Hubbard model, \refqq{Hamiltonian}, can be written as
\begin{eqnarray}
H &=& \sum_{\mathbf{\tilde{k}}, \sigma}\left[\varepsilon^{}_{\mathbf{\tilde{k}}}c_{\mathbf{\tilde{k}}\sigma}^{\dagger}c^{}_{\mathbf{\tilde{k}}\sigma}+\varepsilon^{}_{\mathbf{\tilde{k}}+\mathbf{Q}}c_{\mathbf{\tilde{k}}\sigma}^{\dagger}c^{}_{\mathbf{\tilde{k}}+\mathbf{Q}\sigma}\right] \nonumber\\
&&+ 
U\sum_{\mathbf{k},\mathbf{k^{\prime}}}c_{\mathbf{k}\uparrow}^{\dagger}c^{}_{\mathbf{k}+\mathbf{Q}\uparrow}c_{\mathbf{k^{\prime}}\downarrow}^{\dagger}c^{}_{\mathbf{k^{\prime}}+\mathbf{Q}\downarrow}\:,
\end{eqnarray}
where the sum over $\mathbf{\tilde{k}}$ is restricted to the MBZ, whereas the sum over $\mathbf{k}$ is in the original BZ. After defining a mean-field order parameter for the SDW
\begin{equation}
\Delta = U\sum_{\mathbf{k}} \sigma c_{\mathbf{k}\sigma}^{\dagger}c^{}_{\mathbf{k}+\mathbf{Q}\sigma}\:,
\end{equation}
the mean-field Hamiltonian can be written as
\begin{eqnarray}
H &=& \sum_{\mathbf{\tilde{k}}, \sigma}\left[\varepsilon^{}_{\mathbf{\tilde{k}}}c_{\mathbf{\tilde{k}}\sigma}^{\dagger}c^{}_{\mathbf{\tilde{k}}\sigma}+\varepsilon^{}_{\mathbf{\tilde{k}}+\mathbf{Q}}c_{\mathbf{\tilde{k}}\sigma}^{\dagger}c^{}_{\mathbf{\tilde{k}}+\mathbf{Q}\sigma}\right] \nonumber\\
&& - \Delta\sum_{\mathbf{k}}\left[c_{\mathbf{k}\uparrow}^{\dagger}c^{}_{\mathbf{k}+\mathbf{Q}\uparrow} - c_{\mathbf{k}+\mathbf{Q}\downarrow}^{\dagger}c^{}_{\mathbf{k}\downarrow}\right]\;.
\end{eqnarray}
If we restrict the sum over $\mathbf{k}$ in the second term to be also inside the MBZ and consider only one spin component, we have the following compact form of the mean-field Hamiltonian,
\begin{equation}
H=\sum_{\mathbf{\tilde{k}}}\left(c_{\mathbf{\tilde{k}}}^{\dagger}, c_{\mathbf{\tilde{k}}+\mathbf{Q}}^{\dagger}\right)\begin{pmatrix}
\varepsilon_{\mathbf{\tilde{k}}} & -\Delta \cr
-\Delta & \varepsilon_{\mathbf{\tilde{k}}+\mathbf{Q}}
\end{pmatrix}
\begin{pmatrix}
c_{\mathbf{\tilde{k}}} \cr
c_{\mathbf{\tilde{k}}+\mathbf{Q}}
\end{pmatrix}\;,
\end{equation}
from which the single-particle Green's function can be easily calculated as
\begin{eqnarray}
G_{\mathbf{\tilde{k}}, \omega} &=& \frac{\omega-\varepsilon_{\mathbf{\tilde{k}}+\mathbf{Q}}}
{(\omega-\varepsilon_{\mathbf{\tilde{k}}})(\omega-\varepsilon_{\mathbf{\tilde{k}}+\mathbf{Q}})-\Delta^{2}}\nonumber\\
&=&\frac{1}{\omega-\varepsilon_{\mathbf{\tilde{k}}}-\frac{\Delta^{2}}{\omega-\varepsilon_{\mathbf{\tilde{k}}+\mathbf{Q}}}}\;.
\end{eqnarray}
Thus, the self-energy of the Hubbard model from the SDW mean-field theory is 
\begin{equation}
\label{eq:selfenergy_SDW}
\Sigma(\mathbf{\tilde{k}},\omega) = \frac{\Delta^{2}}{\omega-\varepsilon_{\mathbf{\tilde{k}}+\mathbf{Q}}}=\frac{\Delta^{2}}{\omega+\varepsilon_{\mathbf{\tilde{k}}}}\;, 
\end{equation}
which leads to the strong negative linear-dependence  $\omega = -\varepsilon_{\mathbf{k}}$ at low energies.
Since second-order perturbation theory does not include the magnetic correlations of the system, it is not surprising that, at low temperature, the self-energy calculated from it does not  contain such a negative linear $\varepsilon_{\mathbf{\tilde{k}}}$-dependence. 
We want to note that, despite the simple form of our model function in \refq{modelFct}, it correctly describes the magnetic correlations which only appear at higher orders of perturbation theory. 
Our model function can then be used to describe the competition between the charge and the spin degrees of freedom which becomes important when the temperature decreases.

In \refq{selfenergy_SDW}, we have used the fact that $\varepsilon_{\mathbf{\tilde{k}}+\mathbf{Q}}=-\varepsilon_{\mathbf{\tilde{k}}}$, which holds for a square lattice where $\mathbf{Q}\approx(\pi,\pi)$. For general lattices, $\varepsilon_{\mathbf{k}+\mathbf{Q}}$ is not uniquely related to $\varepsilon_{\vk}$. One might expect that this leads to a somewhat more complicated self-energy parametrization: $\Sigma(\vk, \omega)\to\Sigma(\varepsilon_\vk,\varepsilon_{\vk+\mathbf{Q}}, \omega)$.


\subsection{\protect\boldmath Comparison to D$\Gamma$A}
\label{subsec:CollapseDGA}

\begin{figure}[t] 
	\includegraphics[width=\columnwidth]{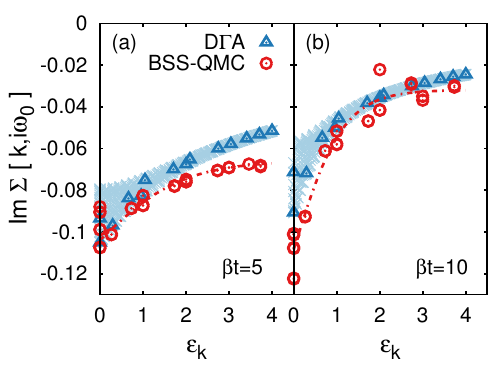}
	\caption{(Color online): Imaginary part of the self-energy $\Sigma(\vk,i\omega_0)$ for $U=2t$ and two different temperatures: (a) $\beta t=5$  and (b)  $\beta t=10$. The (light blue) continuum of data points  represent all different $\DGA$ momenta for a given $\varepsilon_{\mathbf{k}}$. For a better comparison of $\DGA$ with BSS-QMC, we highlighted data points in the $\DGA$ calculation (dark-blue triangles) that correspond to the BSS-QMC data (red circles) with similar $\vk$-points.}
	\label{fig:DGA_tempComp}
\end{figure}

The self-energy parametrization proposed in \refq{modelFct} is based on the large-scale simulation of the BSS-QMC. By comparing the results for different sizes of clusters, we have shown that the 
collapses of the self-energy onto $\Sigma(\varepsilon_\vk,\omega)$ in BSS-QMC 
are observed for all sufficiently large cluster sizes.
In this section, we further confirm that our parametrization of the self-energy via the 
non-interacting dispersion also holds in the thermodynamic limit (including spatial correlations on every length scale). 
Toward that end, we perform calculations by means of the D$\Gamma$A (introduced in \refss{dgamma}). 

\reff{DGA_tempComp} presents the imaginary part of $\Sigma(\varepsilon_{\vk},i\omega_{0})$ for $U=2t$.
The light blue data points in the background correspond to the data at all available $\vk$-points of the D$\Gamma$A calculation. 
Due to a higher resolution in $\mathbf{k}$-space for the $\DGA$ self-energy, we have more $\vk$-points within this method. Those $\vk$-points, similar to those of the BSS-QMC calculation (dotted circles), are represented by dark-blue triangles.
The absolute deviations of the two methods might be due to diagrams beyond ladder D$\Gamma$A or due to the BSS-QMC coarse graining. 
Nevertheless, the collapse of $\Sigma(\varepsilon_{\vk},i\omega_{0})$ away from the Fermi energy $\varepsilon_{\vk}=0$ survives in the thermodynamic limit.
That is, the light blue triangles essentially collapse onto a single line with only minor deviations.
In particular, the behavior of $\Sigma(\varepsilon_{\vk},i\omega_{0})$ in D$\Gamma$A resembles that of BSS-QMC 
in view of the fact that leaving the Fermi edge $\varepsilon_{\vk}=0$, the spread of the data points gets drastically diminished. Again, the relatively big spread at the Fermi edge of the D$\Gamma$A self-energy can be explained physically by the onset of the opening of a pseudogap in this region of the (D$\Gamma$A) 
phase diagram\cite{Katanin2009,Schaefer_Fate2015,Schaefer_Proc2015}. For $U=2t$, the pseudogap phase is relatively small from $T=0.05t$ to $T=0.07t$ (in our units). 
That is, at $T=\frac{1}{\beta}=\frac{t}{5}=0.2t$ the pseudogap has not opened yet but there is already a large scattering; Im~$\Sigma$ at $(\pi,0)$ is already large.\cite{Schaefer_Proc2015}
Reducing the temperature to $T=0.1t$, the spread at the Fermi level increases further. However, the collapse away from the Fermi edge persists [see \reffb{DGA_tempComp}].

Leaving the vicinity of the Fermi edge by choosing a finite Matsubara frequency, the collapse becomes even more drastic as can be seen in the first row of \reff{DGA_freqComp},
where the imaginary part of $\Sigma(\varepsilon_{\vk},i\omega_{n})$ is plotted for $U=4t$ and $\beta{t}=2$ for the first [\reffa{DGA_freqComp}] and second [\reffb{DGA_freqComp}] Matsubara frequency respectively.
One can observe that especially for $\varepsilon_{\vk}\rightarrow{0}$, the spread of $\text{Im }\Sigma(\varepsilon_{\vk},i\omega_{1})$ is much smaller than that of $\text{Im }\Sigma(\varepsilon_{\vk},i\omega_{0})$, again 
a feature signaling the onset of the pseudogap phase.~\footnote{The actually opening of the pseudogap is at $T=0.2t$, at $T=0.1t$ also $(\pi/2,\pi/2)$ becomes gapped for $U=4t$ [see Ref.~\onlinecite{Schaefer_Proc2015}]}
Additionally, the collapse of the 
real part of the self-energy [see the lower row of \reffb{DGA_freqComp}] supports the significance of the energy-parametrization of the self-energy.

Concluding this comparison, the switch from the finite-size cluster in BSS-QMC to 
the thermodynamic limit in D$\Gamma$A does not seem to have any influence on the qualitative phenomenon of the collapse of the self-energy parametrization.

\begin{figure}[htbp] 
	\includegraphics[width=\columnwidth]{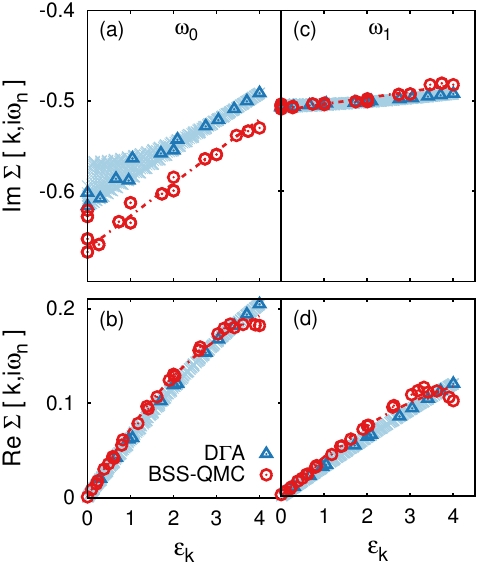}
	\caption{(Color online): The self-energy for $U=4t$, $\beta t=2$ and the first (left) and second (right) Matsubara frequencies comparing BSS-QMC (red circles) and
  $\DGA$ (light blue; and dark blue triangles for similar momenta as the red circles).}
 	\label{fig:DGA_freqComp}
\end{figure}

\subsection{Anisotropic Case}
\label{subsec:AnisotropicCase}

\begin{figure}[htbp]
	\includegraphics[width=\columnwidth]{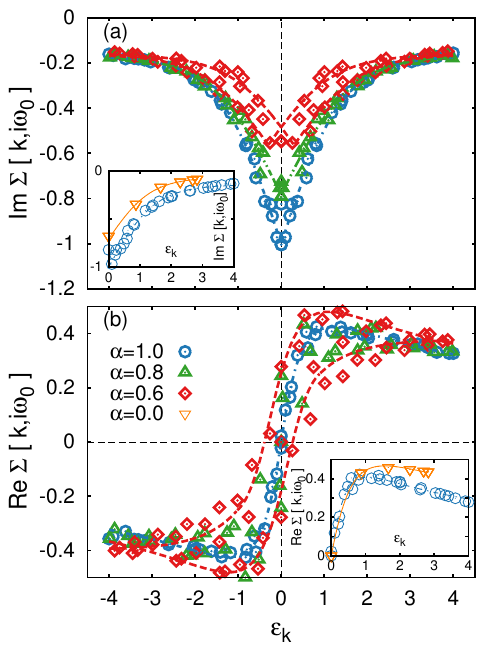}
	\caption{(Color online): Imaginary (a) and real (b) part of the self-energy $\Sigma(\vk,i\omega_0)$ from BSS-QMC at $U = 4t$ and $\beta t= 5.6$ but for various degrees of anisotropies $\alpha$. Green and red lines are guides to the eye only, not fits to the parametrization model as in the other figures.}
	\label{fig:anisotropicCase}
\end{figure}

So far, we have considered the Hubbard model on an isotropic lattice. We found that the two momentum degrees of freedom appearing  (besides the frequency) as variables of the self-energy can be replaced by one energy-like variable with good accuracy and for almost all $\vk$: $\Sigma(\vk,\omega) \equiv \Sigma(\varepsilon_\vk,\omega)$. It is easy to see that such an replacement would be exact (globally) in one dimension (for nearest neighbor hopping): then,
there is only one momentum variable $k_x$. Since $\varepsilon_{k_x}$ increases monotonously with  $k_x\in [0, \pi]$ (in the case of hopping only between nearest neighbors), and $\pm k_x$ are equivalent by symmetry, there exists a unique mapping $\vk \leftrightarrow \varepsilon_\vk$ in one dimension (within the reduced BZ).


The question to be addressed in this section is, whether the parametrization of $\Sigma$ via $\varepsilon_\vk$ works also in the crossover region between these limits. For this purpose, we consider the anisotropic two-dimensional lattice with a hopping ratio $0\le \alpha = t_x/t_y \le 1$; in order to keep the kinetic energy scale $(2t_x^2+2t_y^2)^{1/2}=2t$ fixed, we set $t_y=\sqrt{2t^2/(\alpha^2+1)}$ (and  $t_x=\alpha\, t_y$).

Corresponding BSS-QMC results are shown for $\alpha=1$ (the isotropic case considered before), $\alpha=0.8$, and $\alpha=0.6$ in the main panels of \reff{anisotropicCase}. It is immediately seen that the spread of each data set, associated with an incomplete collapse, increases rapidly with increasing anisotropy, both in the real and imaginary parts of the self-energy. Only in the one-dimensional limit ($\alpha=0$), shown in the insets, do the data fall, again, onto single curves (which are remarkably similar to their two-dimensional counterparts).

Note that Im~$\Sigma$ still shows a reasonably good collapse at $\alpha=0.8$ [triangles in \reffa{anisotropicCase}], while the deviations from a common curve are nearly an order of magnitude larger for Re~$\Sigma$ [triangles in \reffb{anisotropicCase}]. This distinction already hints at the physical reason why a parametrization of the self-energy in terms of the free dispersion cannot work in full generality: In the absence of sufficient symmetries, interactions modify the Fermi surface (while keeping its volume constant at least in the Fermi liquid regime). This direction-dependent shift is encoded, to first order, in Re~$\Sigma(\vk,\omega)\large|_{\varepsilon_\vk=0, \omega=0}$, which would vanish exactly in a parametrization via $\varepsilon_\vk$.

Thus, the analysis of this paper seems to apply directly only to the case of very weak (or very strong) anisotropies. It remains to be seen whether the results of a parametrization such as that performed in \refss{physics} could be useful also in the cases in which the true self-energy does not have this form (as for $\alpha=0.6$) or if the analysis can be extended in order to incorporate Fermi surface deformation.

\subsection{Doping}
\label{subsec:Doping}

\begin{figure}[b]
	\includegraphics[width=\columnwidth]{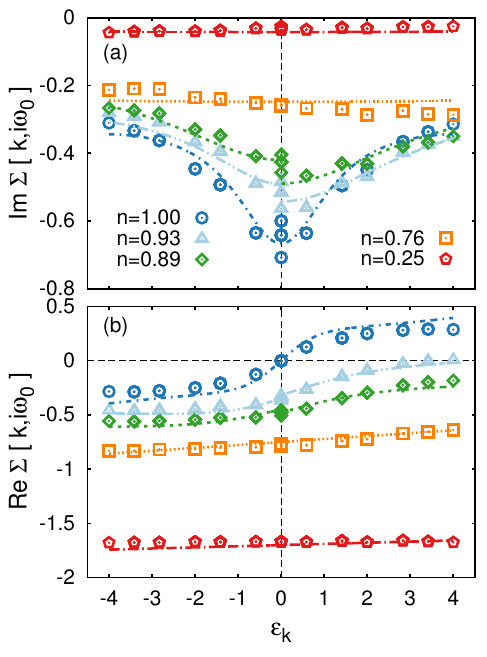}
	\caption{(Color online): Imaginary (a) and real (b) part of the self-energy $\Sigma(\vk,i\omega_0)$ from BSS-QMC at $U=4t$, $\beta t=3.6$ and $L=8\times8$ for doped systems.}
	\label{fig:beyondHalfFilling}
\end{figure}
So far, we have considered the Hubbard model at half-filling ($n=1$). 
Similar insights with respect to the structure of the self-energy would be even more welcome for doped systems, as these are directly relevant for high-temperature superconductivity, {\it i.e.}, physically even more interesting, and also particularly challenging.
However, as BSS-QMC simulations then suffer from the notorious ``minus-sign'' problem, due to the lack of particle-hole symmetry, the numerical effort is much greater (at fixed statistical error).
Consequently, we need to reduce the lattice size  to $8\times8$ in our calculations. 
We should stress, as a caveat, that the resulting reduction in the number of inequivalent $\vk$ points  implies a much sparser $\varepsilon$ grid which makes it more difficult to check for a collapse of $\Sigma$ versus $\varepsilon_\vk$.

Away from half-filling ($n\ne 1$), the self-energy becomes asymmetric with respect to $\varepsilon_{\vk}=0$, as shown in \reff{beyondHalfFilling} for the isotropic 2D case. 
In this figure, we show the (a) imaginary and the (b) real part of the self-energy at five different doping levels characterized by the different values of the electronic density $n$.
Symbols in \reff{beyondHalfFilling} correspond to the BSS-QMC data, and the dotted and dashed lines are obtained by fitting these data with model A in \reft{ModelParameters}.

The first observation for the doped case is that the spread of the self-energy (in the imaginary part) remains at $\varepsilon_{\vk}=0$  and quickly disappears by increasing doping. 
Thus, for a given doping level the self-energy again collapses onto a single curve, which makes a parametrization  possible, as in the half-filled case. 
Note that in order to fit the data in \reff{beyondHalfFilling}, in addition to model A, we added a
constant (with imaginary and real parts) and we took a different  fit model for  $\varepsilon_\vk>0$ and $\varepsilon_\vk<0$.
This way, model~A still  nicely describes the curvature of the  BSS-QMC self-energy. 

For the doped case, we observe deviations from model A [in its original form presented in \refq{modelFct}], but the general form $\Sigma(\vk,\omega) \to \Sigma(\varepsilon_\vk,\omega)$ still holds. 
Again (as in the anisotropic case), the deviations from a smooth dependence of $\Sigma$ on $\varepsilon_\vk$ can be understood as resulting from deformations of the Fermi surface. 

\section{Conclusion \& Outlook}
\label{sec:ConclusionAndOutlook}

Despite the fundamental importance of the self-energy $\Sigma(\vk,\omega)$ within the Hubbard model, little was known about its momentum-frequency structure in the most interesting and challenging cases of $d=2$ and $d=3$ spatial dimensions. One complicating factor in earlier analyses was certainly the high dimensionality $(d+1)$ of the momentum-frequency parameter space, making a full global visualization impossible already in two spatial dimensions.

This situation is changed by our finding that the momentum dependence of the self-energy reduces, with remarkably high precision and scope, to a dependence on the non-interacting energy $\varepsilon_{\vk}$ at each point in momentum space, {\it i.e.} $\Sigma(\vk,\omega) \to \Sigma(\varepsilon,\omega)$ on a square lattice, where $\varepsilon=\varepsilon_\vk$. Thereby, we could not only fully visualize the numerically obtained self-energy in the density plots of \reffa{Sigma_GF_B056} and \reffa{Sigma_GF_B020} at temperatures in and above the pseudogap phase, respectively (note that these spectral data also determine Re~$\Sigma$), but we could also derive complete parametrizations that highlight the interesting physics previously hidden in this system. We could trace back the strong \inverseZ shaped low-$T$ structure to the generation of (self-energy) spectral density  at $\omega=\varepsilon_{\vk+\vQ}=-\varepsilon_\vk$ by antiferromagnetic fluctuations. For other lattices $\varepsilon_{\vk+\vQ}\neq-\varepsilon_\vk$, suggesting a parametrization $\Sigma(\varepsilon_\vk,\varepsilon_{\vk+\vQ},\omega)$.

Given this explanation, one might have expected the spectral features to decay only weakly towards higher temperatures, similarly to the nearest-neighbor spin correlation function. However, the higher-$T$ results completely lack any (lower-energy) features at $\omega=-\varepsilon_\vk$, and they show, instead, significant contributions at $\omega=\varepsilon_\vk$, leading to an overall \script{Z}-shaped structure that appears also in second-order perturbation theory.

Note that our ansatz for the self-energy is the most general one consistent with the functional form of the Green's function $G\equiv G(\varepsilon_\vk,\omega)$ that is valid also within DMFT. However, it is clear that DMFT taps only a very limited subspace of this class of Green's functions.

Limitations of the ansatz $\Sigma \equiv \Sigma(\varepsilon_\vk,\omega)$ become apparent both directly at the Fermi surface in the pseudogap phase and, more globally, in the case of strongly anisotropic lattices. In the former case, the breakdown is inevitable, since an anisotropic gap cannot possibly be described by a self-energy that is constant along the Fermi surface (at each fixed frequency). This also holds at temperatures somewhat above the pseudogap phase, where the scattering rates at the nodal and antinodal point of the Fermi surface are very different. In the latter case, the physics behind the deviation is the deformation of the Fermi surface. 
For the doped square lattice the general form $\Sigma(\vk,\omega) \to \Sigma(\varepsilon_\vk,\omega)$ is still applicable, albeit our model parametrization does not work any longer.

Another fascinating feature of our ansatz is that it allows for a direct comparison of self-energies associated with systems of different spatial dimensionality (such as those shown in the insets of \reff{anisotropicCase}), as the parameter space is always two-dimensional. In fact, it is reasonable to assume that the analysis of this paper would work even better (with even greater reductions of the complexity) for cubic lattices, {\it i.e.} in three dimensions. However, a reliable verification would require quite large lattices (at still high numerical precision) and it was, therefore, beyond the scope of this work.

\section{acknowledgment}

We acknowledge financial support through the DFG research unit FOR 1346 (PP, DR, NB) and the European Research Council under the European Union's Seventh Framework Program (FP/2007-2013)/ERC through grant agreement n.\ 306447 (PP, GL, KH); and support from the Austrian Science Fund (FWF) through the doctoral school ``Building Solids for Function'' (FWF W1243) and project I-610 (TS).
Calculations have been done in part on the Vienna Scientific Cluster (VSC). 


\bibliography{ref}

\begin{thebibliography}{71}
\bibitem [{Sak()}]{Sakairef}%
  \BibitemOpen
  \href@noop {} {}\bibinfo {note} {Such a \protect\inverseZ-shaped structure
  can also be identified in Fig.~1(a) of
  Ref.~\onlinecite{PhysRevLett.102.056404} showing the zeros of the $G$. Note
  that a maximum in $\Sigma$ corresponds to a minimum (zero) in $G$. The
  high-energy zeros of $G$ are split in this figure into multiple poles because
  of finite-size effects. In Ref.~\onlinecite{PhysRevLett.116.057003}, the
  low-energy shape is modeled by hidden Fermions}\BibitemShut {NoStop}%
\end{thebibliography}

\end{document}